\documentstyle[aps,prl,twocolumn,epsfig,floats]{revtex}

\begin{document}

\bibliographystyle{prsty}

\draft 

\title{Changing the Electronic Spectrum of a Quantum Dot by Adding
Electrons}

\author{S. R. Patel, D. R. Stewart, and C. M. Marcus}
\address{Department of Physics, Stanford University, Stanford, California
94305}

\author{M. G\"{o}k\c{c}eda\u{g}, Y. Alhassid, and A. D. Stone}
\address{Center for Theoretical Physics, Sloane Physics Laboratory,
 Yale University, New Haven, CT 06520}

\author{C. I. Duru\"{o}z and J. S. Harris, Jr.}
\address{Electrical Engineering Department, Stanford University, Stanford,
California 94305}

\date{\today}
\maketitle

\begin{abstract} The temperature dependence of Coulomb blockade peak height
correlation is used to investigate how adding electrons to a quantum dot
alters or ``scrambles" its electronic spectrum. Deviations from
finite-temperature random matrix theory
 with an unchanging spectrum indicate spectral scrambling after a small
number of electrons are added.  Enhanced peak-to-peak correlations at low
temperature are observed. Peak height statistics show similar behavior in
several dot configurations  despite significant differences in correlations.
\end{abstract}


Electron transport through irregular quantum dots -- i.e. micron-scale
islands of confined charge weakly connected to electronic reservoirs -- are
expected to, and in
 some cases actually do, exhibit universal statistics associated with
quantum chaos
\cite{Reviews}. An example where theory and experiment  agree well  is the
distribution of Coulomb blockade (CB) peak heights
\cite{Jalabert92,Prigodin93,Chang96,Folk96}. At temperatures $T$ that are
much
 smaller than the mean level spacing of the dot $\Delta$, transport on a CB
peak is mediated by resonant tunneling through a single level
 \cite{Beenakker91,KouwenhovenReview}. Large fluctuations in CB peak heights
in this regime reflect the fluctuating strength of coupling of the chaotic
wavefunction in the dot to the modes in the leads, leading to universal
statistics sensitive only to time-reversal symmetry
\cite{Jalabert92,Prigodin93}, in good agreement with experiment
\cite{Chang96,Folk96}.

At higher temperatures,
 $\Delta < k_B T < E_C$, where $E_C$ is the classical charging energy, each
CB peak contains contributions from $\sim k_B T /\Delta$ quantum levels, and
one would expect roughly this number of adjacent peaks to be correlated in
height. This assumes that the spectrum of the dot  does not change as
electrons are added.
 On the other hand, if adding electrons alters
 the  spectrum, then the correlation length in peak number, $n_c$, will not
grow beyond a certain value,
$m$, which roughly measures (but is not equivalent to) the number of added
electrons needed to completely ``scramble" the electronic spectrum.

This Letter presents measurements of the temperature dependence of the CB
peak-to-peak height correlation and peak height statistics for
 gate-confined GaAs quantum dots, and compares these results to finite
temperature random matrix theory (RMT) calculations that neglect spectral
scrambling \cite{Alhassid98}.    We find that the  number of correlated
peaks $n_c(T)$ saturates at $m \sim 2-5$, with smaller dots saturating at
smaller $m$. At the low temperature end, we find that
$n_c(T)$ is {\it larger} than the value predicted by RMT.  That is,
correlations in peak heights beyond thermal smearing exist for reasons that
are not clear. Some possible explanations are considered below.

In contrast to the dependence on dot configuration found for the peak height
correlations (as reflected in $n_c(T)$ and $m$), peak height {\it
statistics} are found to be very similar for all device configurations.
This suggests that peak statistics are more robustly ``universal''  than
peak correlations, not surprising considering that distributions are not
sensitive to spectral scrambling. The ratio of the
 standard deviation to mean of peak heights  is found to be
 smaller than predicted, possibly due to the effects of decoherence.

What does one expect to be the effect of adding electrons on the spectrum of
a quantum dot?  In the limit of weak electron-electron interactions (and
neglecting shape deformations caused by changing gate voltages)
 a fixed spectrum of  single-particle states is simply filled one at a time,
leading to  $n_c(T)
\sim k_B T / \Delta$ and $m \gg1$. In the opposite limit of strong
interactions, the spectrum could be totally scrambled with the addition of
each  electron,
 giving $m \sim 1$. For a  GaAs  quantum dot containing many ($\sim 100$ or
more) electrons, RPA calculations
\cite{Blanter97,Berkovits97} (appropriate for weak interactions) indicate
that fluctuations
 in the ground state energy due to interactions are small, of order
$r_s g^{-1/2}\Delta$ where $g$ is the dimensionless conductance of the dot
and  $r_s$ is the so-called gas parameter, the ratio of potential to kinetic
energy of the electrons ($r_s \sim 1-2$ in GaAs heterostructures). For a
ballistic-chaotic dot containing ${\cal N}$ electrons
$g \sim {\cal N}^{1/2}$, giving a rough estimate for the number of electrons
needed to
 scramble the spectrum, $ 1 < m < \sim {\cal N}^{1/2}/r_s^2$, assuming that
fluctuations accumulate
 randomly as electrons are added to the dot.

  Measurements of CB peak  {\it spacing} statistics have in some cases found
 rms fluctuations in $E_C$ as large as 15\% \cite{Sivan96,Simmel97},
consistent with classical estimates \cite{Koulakov97} and numerics
\cite{Sivan96}  for strong interactions (where RPA fails), suggesting that
one or a few electrons can significantly alter the charge arrangement of the
dot. Other experiments
\cite{Patel98} have found smaller fluctuations, of order $\Delta$,
suggesting a lesser  degree of rearrangement. In all experiments the
peak-spacing distribution is found to be roughly  Gaussian, which is
surprising considering that spin degeneracy would naively imply a bimodal
distribution. Suggested explanations include wave-function dependent
interactions
\cite{Blanter97,Stopa97} and scrambling due to shape deformation
\cite{Vallejos98}. Magnetofingerprints of CB peaks in dots with $\sim
50-100$ electrons show a
 persistence of both the addition and excitation spectrum over at least $6$
peaks, with changes in the excitation spectrum (measured by nonlinear
magnetotransport) of order
$\Delta$ upon adding one electron, leading occasionally to an exchange of a
pair of levels in the spectrum \cite{Stewart97}. Symmetric, few-electron
dots also show a sort of scrambling in the sense that the ground state
shell-filling structure may differ from the corresponding excited state
before the ${\cal N}^{th}$ electron is added, for as few as
${\cal N} = 4$ electrons \cite{Tarucha96}.  For such small systems (${\cal
N} < \sim 5-6$) exact calculations are possible, and a statistical approach
to scrambling is not necessary.

Before describing the experiment, we discuss the  generalization of
 the theory of CB peak height fluctuations to temperatures that are
comparable or greater than
$\Delta$, but without including  spectral  scrambling
\cite{Beenakker91,Alhassid98}. Well-formed
 CB peaks require weak tunneling from each of the
 levels $\lambda$ coupling left and right leads to the dot, with tunneling
rates
$(\Gamma_l^\lambda,\Gamma_r^\lambda) \ll \Delta / h$, and also low bias and
temperature,
$(e V_{ds},k_BT) \ll E_C$, where $V_{ds}$ is the voltage bias across the
dot.
In this regime, the peak conductance
$G_{max}$ has the form
\begin{equation} G_{max} = {{e^2} \over {h}} {{h \overline \Gamma}
\over {8 k_{\rm B} T}} \alpha(T),
\label{gmax}
\end{equation} where $\alpha(T) = \sum_\lambda  \alpha_\lambda w_\lambda(T)$
is a
 weighted sum of normalized lead-dot-lead conductances,
$\alpha_\lambda = 2 \Gamma_l^\lambda
\Gamma_r^\lambda / (\overline \Gamma (\Gamma_l^\lambda+\Gamma_r^\lambda))$.
The dot is assumed to be symmetrically coupled to the leads, with average
tunneling rates
$\overline {\Gamma_l^\lambda} = \overline {\Gamma_r^\lambda} = \overline
\Gamma / 2$. In the experimentally relevant regime
$(k_{\rm B} T,
\Delta) \ll E_C$, the thermal weights $w_\lambda( T)$ are given by
  $w_\lambda( T) = 4 f(\Delta {\rm F}_{\cal N} - {\widetilde E}_F) {\langle
n_\lambda \rangle}_{\cal N} [1-f(E_\lambda-{\widetilde E}_F)]$, where
$\Delta {\rm F}_{\cal N}= {\rm F}_{\cal N} - {\rm F}_{{\cal N}-1}$ is the
difference in the canonical free energy of
${\cal N}$
 and ${{\cal N} -1 }$ non-interacting electrons on the dot,
${\langle n_\lambda \rangle}_{\cal N}$ is the canonical  occupation of level
$\lambda$ with ${\cal N}$ electrons on the dot, ${\widetilde E}_F = [{E_F} +
e \eta V_g  - ({\cal N} - 1/2) E_C]$ is
 the effective Fermi energy with
$V_g$ tuned between ${\cal N} -1 $ and ${\cal N}$ electrons on the dot,
 $E_\lambda$ is the energy of level $\lambda$, and $f(\epsilon) =
 1/(1+e^{\epsilon/ k_{\rm B} T})$ is the Fermi function. Equation
(\ref{gmax}) generalizes previous results for low temperatures, $h\overline
\Gamma \ll k_{\rm B} T
\ll
\Delta$, and yields the known distributions for
$\alpha$ in that limit \cite{Jalabert92,Prigodin93}.

Within a noninteracting model, no correlations between neighboring peak
heights are expected for $k_BT \ll \Delta$. At higher temperatures,
correlations appear as each level is able to contribute to several nearby
peaks. For both numerical RMT data and experimental data, we compute a
discrete correlation
 function $C(n)$ from a sequence of $N$ peaks $(G_{max})_i$
\begin{equation} C(n) = {{1 \over {{N}-n}} \sum^{{N}-n}_{i=1}
 {\delta g}_i{\delta g}_{i+n}} \bigg/ {{1 \over {N}} \sum^{N}_{i=1} {\delta
g}_i {\delta g}_i},
\label{cdeltan}
\end{equation} where ${\delta g}_i = ((G_{max})_i - {\langle  G_{max}
\rangle}_{{N},B})$ is the fluctuation of the $i^{th}$ peak height  around
the
average (calculated over both peak number and magnetic field).  The
correlation length
$n_c$ is then calculated from  a Gaussian fit,
$C(n) = e^{-(n/n_c)^2}$. The gaussian form is not based on any theoretical
model but appears to accurately describe the shape of both the
RMT and experimental data.

 The quantum dots we discuss are fabricated using e-beam lithography to
pattern Cr/Au gates on the surface of a GaAs/AlGaAs heterostructure 900
$\rm\AA$  above the 2DEG layer.  Multi-gate dot design allows changes in
device size and shape by changing gate voltages. Active control of point
contact gates during sweeps of the ``plunger" gate $V_g$ compensates any
unintentional capacitive coupling, allowing many peaks to be swept over
without changing the average transmission of the leads. All data were taken
using two-wire lock-in techniques with
$2\ \mu eV$ bias at $13\ Hz$.  An electron base temperature of $~ 45\ mK$
and
ratio $\eta$ of gate voltage to dot energy were extracted from the CB peak
width versus temperature using standard methods
\cite{Beenakker91}.  The charging energy $E_C$ in each configuration
 was calculated using $\eta$ and the average peak spacing in gate voltage
\cite{KouwenhovenReview}. The mean level spacing
$\Delta$ was measured from the sub-structure (corresponding to excited
states)
 in the  differential conductance at finite bias using a small ac signal
added to a dc bias. Measurements were made with broken time-reversal
symmetry, with $3-10\
\phi_o$ through the device area. For each temperature,
$V_g$ was scanned over 50-100 peaks and $B$ was then changed by $\sim
\phi_o/A_{dot}$ to give independent peak height statistics.

\begin{figure}[bth]
\epsfxsize= 3 in \epsfbox{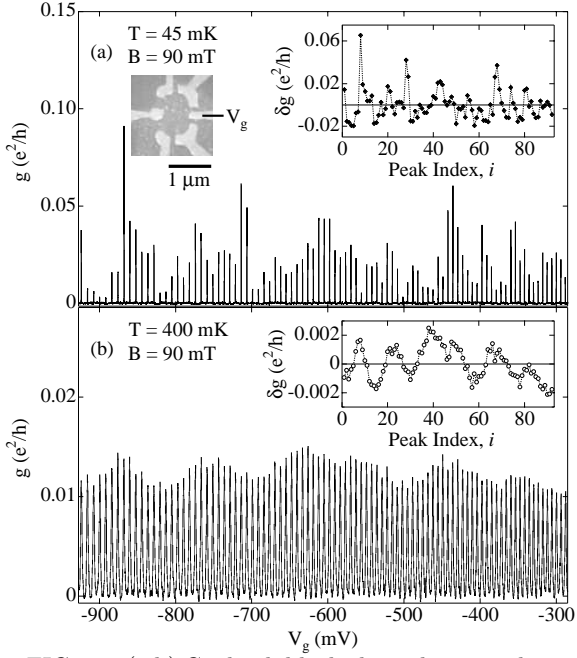}
  \caption{ (a,b) Coulomb blockade peaks in conductance $g$ as a function of
gate voltage $V_g$ at (a)
$45\ mK$  and (b) $400\ mK$ from device 1. Insets: SEM micrograph of device
1.
 Peak height fluctuations $\delta g_i$ extracted from these data sets.}
  \label{fig1}
\end{figure}

\begin{figure}[bth]
\epsfxsize= 3 in \epsfbox{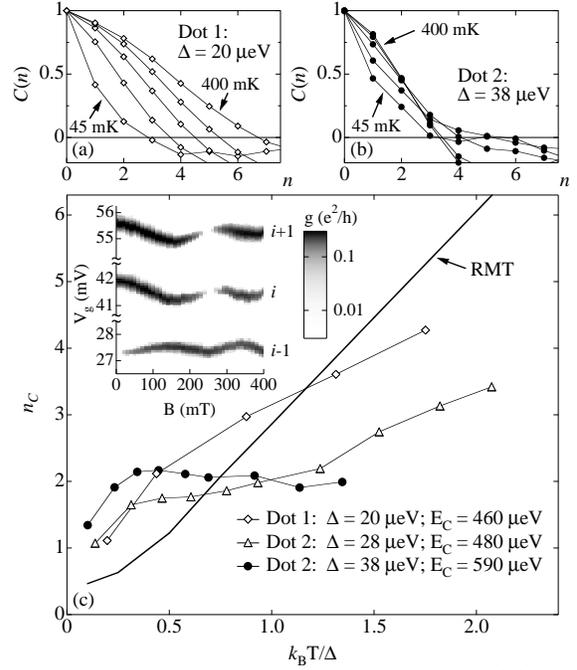}
  \caption{ (a,b) Peak height correlations $C(n)$ at $45\ mK$, $100\ mK$,
$200\ mK$,
$300\ mK$ and $400\ mK$ for (a) dot 1 and (b) dot 2.  (c) Temperature
dependence of correlation length
$n_c$ for different device configurations, and numerical RMT result.
 Inset:  Grayscale plots of conductance for 3 successive CB peaks, showing
paired peaks $i$ and $i+1$, presumably a spin pair.}
  \label{fig2}
\end{figure}

Figure \ref{fig1} shows two typical series of CB peaks at lower and higher
$T$, where  the increased correlation between peaks at higher $T$ is clearly
evident. Peak height fluctuations ${\delta g}_i$ (right insets in Fig.\
\ref{fig1}) are extracted from each series using  fits to
$\cosh^{-2}$ lineshapes around each peak, and the correlations  $C(n)$ are
then calculated according to Eq.\ (\ref{cdeltan}). Plots of $C(n)$ for
 two dots are shown in Figs.\  \ref{fig2}(a, b).  For both configurations,
 the correlation length $n_c$ increases with increasing $T$ at lower $T$.
However, for the smaller configuration  $n_c$ saturates above
$300\ mK$, while for the larger configuration  $n_c$ continues to increase
well above this temperature. In still larger devices, saturation is not
observed up to  temperatures of $700\ mK$.

Correlation lengths $n_c(T)$ for three measured dot configurations ($\Delta
= 20, 28$, and
$38\ \mu\rm eV$) are shown in Fig.\ \ref{fig2}(c) along with the RMT
results. Each data point in Fig.\ \ref{fig2}(c) represents data from $\sim
500$ CB peaks. The RMT curve was computed by  applying Eq. (\ref{cdeltan})
to
a peak sequence that is similar in length to the experimental
 data,  and generated according to  Eq. (\ref{gmax}) assuming  a
uniformly-spaced spectrum \cite{Alhassid98}.  The RMT results do not change
significantly when Wigner-Dyson statistics for
 the spectrum is included.
 The saturation of $n_c(T)$ at $m\sim 2$ for
$k_B T > 0.5\ \Delta$  for the smallest device is  evident in Fig.\
\ref{fig2}(c).  The larger dot begins to saturate for larger $n$,
 with a larger ratio
$k_BT/\Delta$, suggesting that the spectrum of the larger dot is less prone
to scrambling. The observed scale of saturation, $m$, as well as the trend
for $m$ to increase with ${\cal N}$, appears consistent with the RPA
estimate given above.

\begin{figure}[bth]
\epsfxsize=3 in \epsfbox{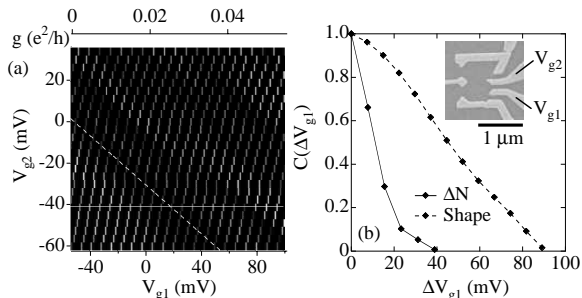}
  \caption{ (a) Grayscale conductance plot of CB peaks as a function
 of $V_{g1}$ and $V_{g2}$ from device shown in inset of (b), with
$\Delta \sim 23\ \mu \rm eV$, at $90\ \rm mK$. The appearance of peaks as
short vertical bars reflects the coarser sampling of $V_{g2}$ compared to
$V_{g1}$; the patterns of connected bars moving up and to the right are
an artifact of this display. (b) Correlation function
$C(\Delta V_{g1})$ of peak height fluctuations for fixed ${\cal N}$ (dashed
curve) and fixed
$V_{g2}$ (changing ${\cal N}$, solid  curve).}
  \label{fig3}
\end{figure}

As the gate voltage is swept, two distinct changes occur,  both of
 which can cause spectral scrambling.  The first is that the  number of
electrons and size of the dot change; the second is that the shape of the
dot
changes due to local movement of the boundary at the position of the gate.
This second effect was considered recently in Ref.
\cite{Vallejos98} to explain the nearly-Gaussian peak spacing distribution
seen in several experiments \cite{Sivan96,Simmel97,Patel98}. The two effects
can be separated using a dot with {\it two} plunger gates, which allows pure
shape distortion without changing  ${\cal N}$ by increasing one gate voltage
and decreasing the other. In practice, it is easier to raster over the two
gate voltages, as seen in Fig.\ \ref{fig3}(a). Horizontal and vertical
directions correspond to single-gate CB measurements, while a downward
diagonal following a single peak corresponds to pure shape distortion with
fixed ${\cal N}$. Correlations in the same dot
 measured at fixed
${\cal N}$ (measured along diagonals) and changing ${\cal N}$ (measured
along
horizontals) can be compared by evaluating both correlations in terms of
$V_{g1}$ rather than $n$. Comparing $C(\Delta V_{g1})$ for the two cases
(Fig.\ \ref{fig3}(b)) shows that the correlation length associated with
shape
deformation is larger by a factor of
$\sim 4$ than that associated with a changing
${\cal N}$.  This indicates that the saturation of $n_c$ (scrambling) is
dominated by changes in electron number rather than by shape distortion.
Further work
 is needed to determine if this result is universally true, but it appears
to
hold in a variety of gate-confined dots that we have measured.

All dot configurations show an enhanced correlation length
 $n_c(T)$ at low $T$ compared to RMT, as seen in Fig.\
\ref{fig2}(c).   Thus the data suggest that temperature alone does not
explain the enhancement, at least within RMT. Several possible explanations
for this have been proposed
\cite{Stopa97,Hackenbroich97,Baltin98}; we further note that correlations at
low
$T$ can result from similar peak heights in spin-paired levels (see Fig.\
\ref{fig2}(c), inset) which may appear in either adjacent or non-adjacent
peaks, depending on the size of interaction-induced spin splitting.

\begin{figure}[bth]
\epsfxsize=2.8 in \epsfbox{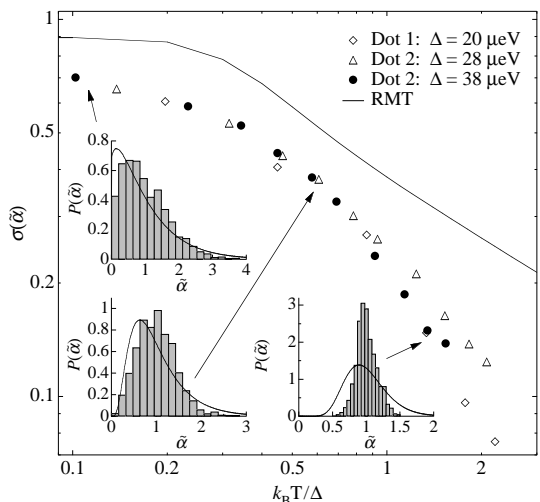}
  \caption{ Temperature dependence of scaled peak height distribution width,
$\sigma(\tilde\alpha)$, along with numerical RMT result (solid curve).
Insets:
 Full distributions of scaled peak heights  $P(\tilde\alpha)$ for all dot
configurations combined, at
$k_{\rm B} T = 0.1\ \Delta$, $k_{\rm B} T = 0.5\ \Delta$, and $k_{\rm B} T =
1.5\
\Delta$, along with RMT distributions. RMT results use a uniformly-spaced
spectrum (rather than Wigner-Dyson).}
  \label{fig4}
\end{figure}

Finally, we investigate the statistics of peak height fluctuations as a
function of temperature. It is convenient to consider a normalized
distribution to remove any temperature dependence of the average peak
height, $\langle
\alpha(T)\rangle$. We define
$\tilde\alpha = \alpha / \langle \alpha \rangle$, so that the distribution
$P(\tilde\alpha)$ will have
$\langle \tilde\alpha(T)\rangle = 1$ for all $T$. The standard deviation
$\sigma(\tilde\alpha)
 \equiv \sqrt{\langle {\tilde\alpha}^2 \rangle -1}$, which characterizes the
width of the distribution relative to its mean (i.e.,
$\sigma(\tilde\alpha)=\sigma(\alpha)/\langle
\alpha \rangle$), is shown for three dot configurations along with RMT
results in Fig.\
\ref{fig4}. All experimental data show very similar temperature  dependences
despite significant differences in correlations described above. We conclude
that peak height distributions, which are {\it not} sensitive to scrambling,
show
 more ``universal" behavior than correlations. Notice, however, that the
experimental data all have smaller height fluctuations than predicted by
RMT.  This can also be seen in the full distributions
$P(\tilde\alpha)$ comparing experiment (histograms) and RMT (solid lines),
shown as insets in Fig.~\ref{fig4}. The departure from finite-temperature
RMT is likely due to decoherence effects, and might provide a novel tool for
measuring decoherence in nearly isolated structures. Ongoing work in this
direction is in progress.

We thank A. G. Huibers and S. M. Grossman for contributions to the
experiment, and O. Agam, I. Aleiner, B. Altshuler, Y. Gefen, I.V. Lerner, B.
Muzykanskii, and M. Stopa
 for valuable discussions. Work at Stanford supported by the ARO
DAAH04-95-1-0221, ONR-YIP N00014-94-1-0622, the NSF-NYI and PECASE programs
(Marcus Group); and JSEP DAAH04-94-G-0058 (Harris Group).
 Work at Yale was supported in part by the DOE DE-FG-0291-ER-40608
 and the NSF DMR-9215065.

\end{document}